\documentclass[
    ,final            
  ]
  {aipproc}

\layoutstyle{6x9}

\begin{document}

\title{Radio observations of the cool gas, dust, and star formation in the first galaxies}

\classification{98.54}
\keywords {First Stars and Galaxies: Challenges in the Next Decade, AIP, 2010}

\author{C. Carilli (NRAO), F. Walter (MPIA), R. Wang (NRAO/Arizona),
D. Riechers (CIT), J. Wagg (ESO), X. Fan (Arizona), K. Menten (MPIfR),
F. Bertoldi (Bonn), P. Cox (IRAM)}{
  address={}
}

\begin{abstract}

We summarize cm through submm observations of the host
galaxies of $z \sim 6$ quasars. These observations reveal the cool
molecular gas (the fuel for star formation), the warm dust (heated by
star formation), the fine structure line emission (tracing the CNM and
PDRs), and the synchrotron emission. Our results imply active
star formation in $\sim 30\%$ of the host galaxies, 
with star formation rates $\sim 10^3$ M$_\odot$ year$^{-1}$,
and molecular gas masses $\sim 10^{10}$ M$_\odot$.  Imaging of the
[CII] emission from the most distant quasar reveals a 'maximal
starburst disk' on a scale $\sim 1.5$ kpc. Gas dynamical studies suggest
a departure of these galaxies from the low-$z$ M$_{BH}$ -- M$_{bulge}$
relation, with the black holes being, on average, 15 times more
massive than expected. Overall, we are witnessing the co-eval
formation of massive galaxies and supermassive black holes within 1
Gyr of the Big Bang.

\end{abstract}

\maketitle


\subsection{Quasar host galaxies at $z \sim 6$}

Remarkable progress has been made in the study of galaxies back
into cosmic reionization, or 'first new light' in the Universe ($z \ge
6$), at optical through near-IR wavelengths. However, such observations
are limited to the stars and ionized gas, the products of star
formation in galaxies. Recent radio observations have begun to probe
the 'other half' of the story, namely, the cool gas and dust, the fuel
for star formation in galaxies, as well as star formation itself, 
unhindered by obscuration. We summarize these recent observations,
and discuss the dramatic increase in potential with the EVLA and ALMA.

Radio studies of the first galaxies have focused primarily on the host
galaxies of luminous quasars, for a number of (mostly practical)
reasons. First, the number of $z \ge 6$ quasars has increased
dramatically in the last decade through the SDSS and UKIDSS surveys
(Fan et al. 2006, Willott et al. 2010). Second, quasars are the only
population of such extreme $z$ galaxies with spectroscopic redshifts
(with one exception), as required for molecular line follow-up
observations give current bandwidth limits.  And third, quasar
demographics, as well as the black hole -- bulge mass relation, suggest
that these are massive (proto-)galaxies, and current radio
telescopes can only detect the more massive galaxies at such high
redshifts. The black hole masses in these systems are $\sim 10^9$
M$_\odot$, implying bulge masses $\sim 10^{12}$, based on the low-$z$
relation (H\"aring \& Rix 2004).

Study of very early massive galaxy formation has become topical with
the recent evidence suggesting that massive galaxies form most of
their stars early and quickly, and the more massive, the earlier and
quicker. Evidence includes: (i) stellar population studies of nearby
ellipticals, (ii) study of specific star formation rates in galaxies
as a function of redshift, and (iii) observation of 'red and dead'
ellipticals at $z \ge 1.5$. In this context, quasar host galaxies at
$z \sim 6$ represent laboratories for the study of very massive
galaxies at the most extreme redshifts.

\subsection{FIR luminous quasar host galaxies: Dust and star
formation}

We have a long-standing program to study the thermal emission from
warm dust in large samples of quasars from $z \sim 2$ to 6, using
(sub)mm bolometer cameras such as MAMBO (Wang et al. 2009). We find
that about 1/3 of high redshift quasar host galaxies are
hyper-luminous IR galaxies ($L_{FIR} \sim 10^{13}$ L$_\odot$), and
that this fraction is roughly constant with redshift out to $z \sim
6$. The implied dust masses are of order $10^8$ M$_\odot$. In this
paper, we concentrate on the 33 quasars at $z > 5.7$ to 6.4,
comprising all the quasars known at these redshifts as of 2008.

The detection of large dust masses within 1Gyr of the Big Bang
immediately raises an interesting question: how to form so much dust
so early?  The standard dust formation mechanism in the ISM involves
coagulation in the cool winds from low mass stars (Draine 2003),
which, naively would take too long ($\ge 10^{9.3}$ years; although
cf. Venkatesan et al. 2006). The large dust masses have led to a number
of theoretical studies of early dust formation, with models ranging
from dust formation associated with massive star formation in eg.
supernova remnants (Dwek et al. 2007), to dust formation in outflows
from the broad line regions of quasars (Elvis et al. 2002). Recent
observations of the UV-extinction curves in a few $z \sim 6$ quasars
and GRBs suggest a different dust composition at these very high
redshift quasars relative to the Milky Way or the SMA, as well as relative 
to quasars at $z < 4$.  The extinction can be modeled by larger silicate
and amorphous carbon grains (vs. eg. graphite), as might be expected from
dust formed in supernova remnants (Stratta et al. 2007
Perley et al. 2010). The formation of dust in the early Unverse remains an
interesting open question.

\begin{figure}
  \includegraphics[height=.2\textheight]{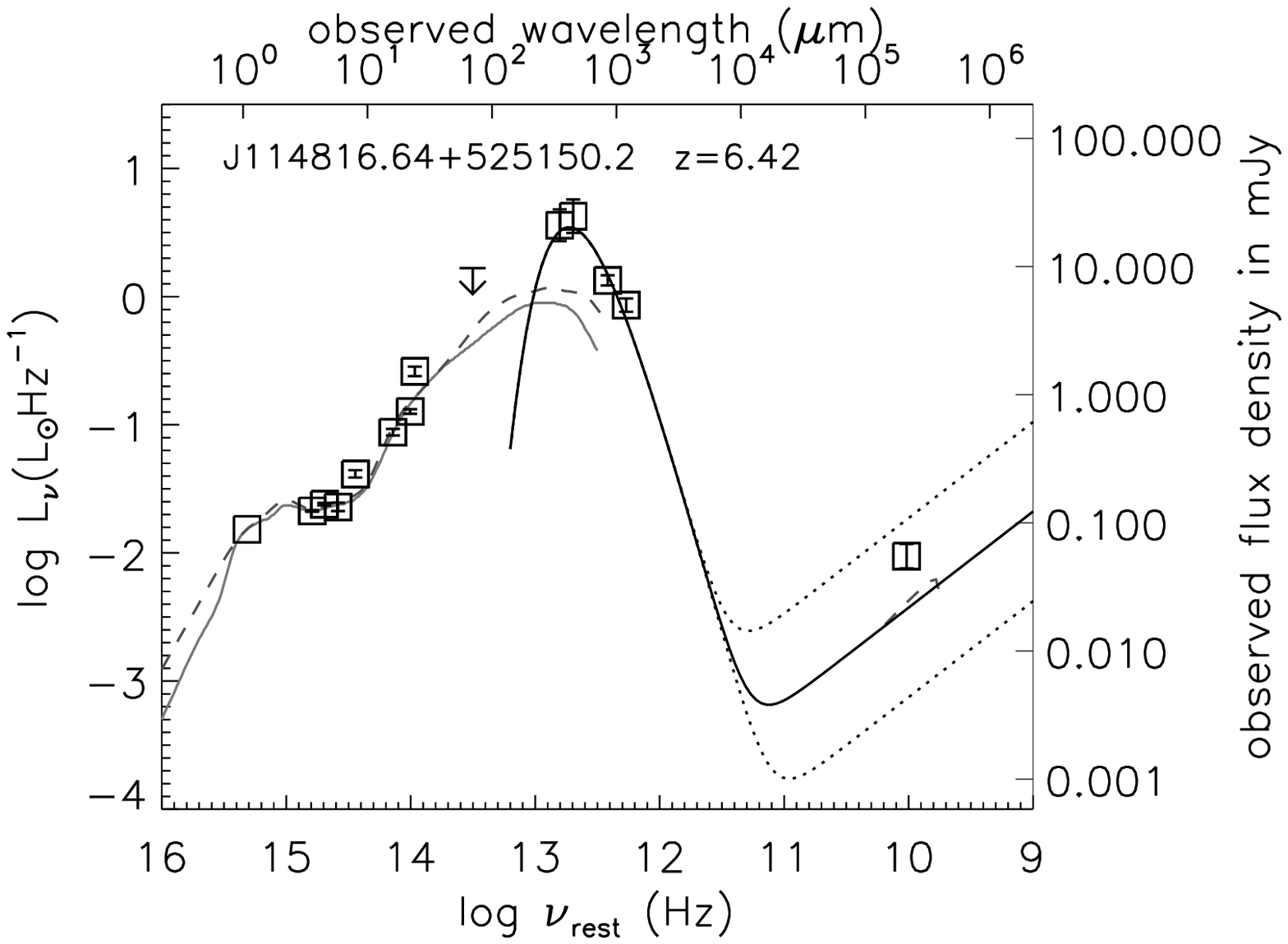}
  \includegraphics[height=.2\textheight]{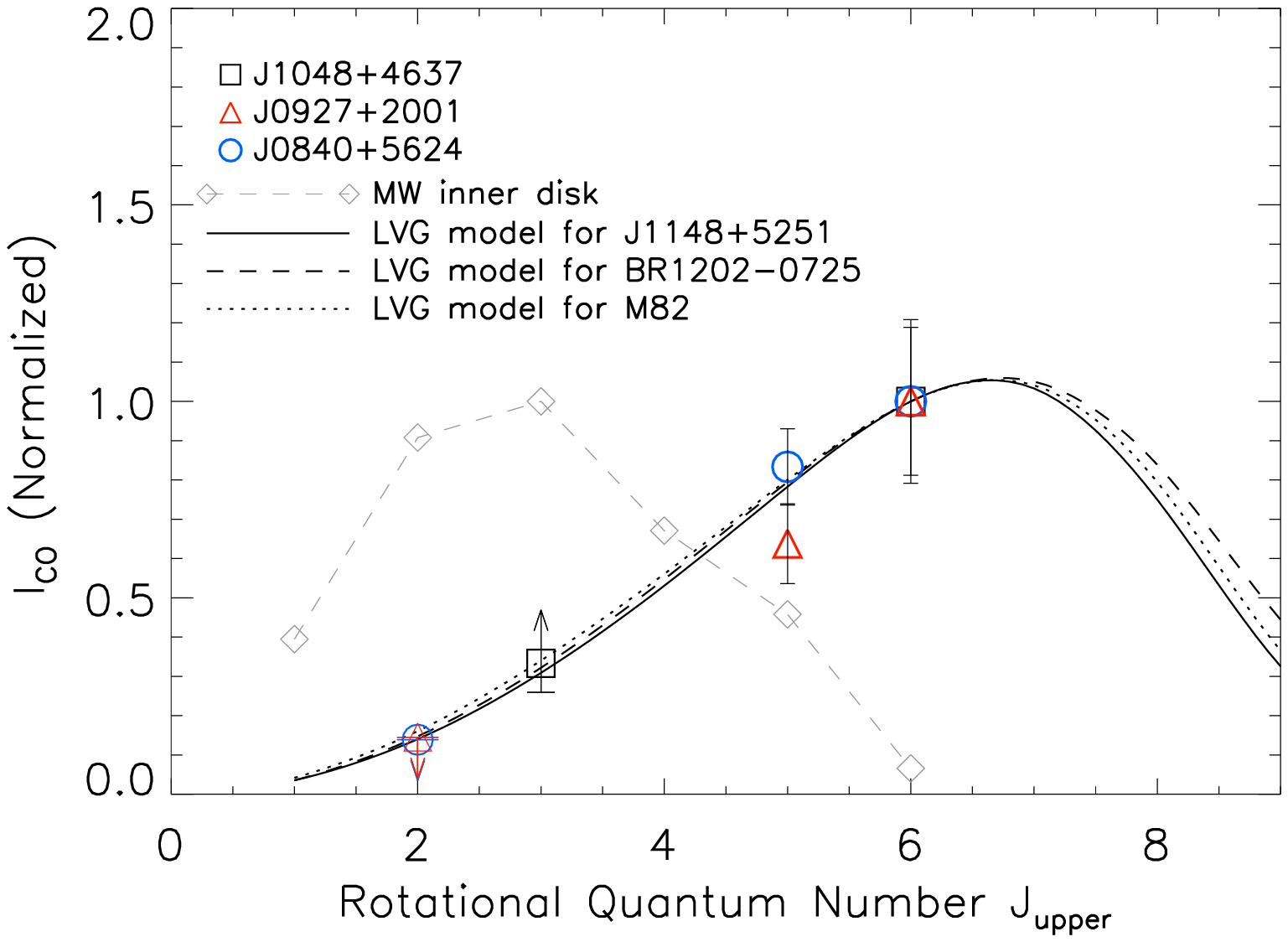}
  \includegraphics[height=.2\textheight]{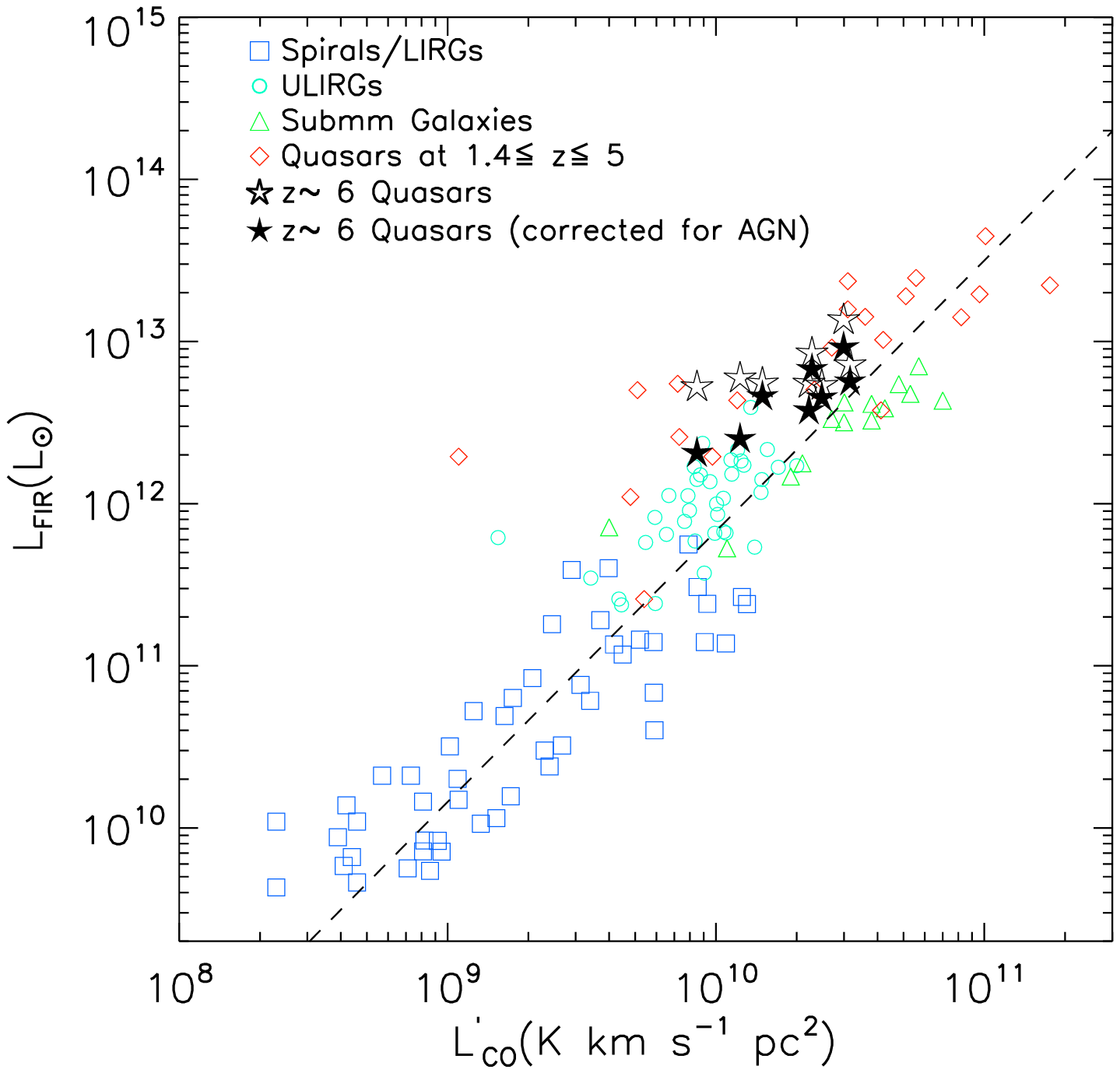}
  \caption{Left: The SED for J1148+5251 at $z =6.42$ from Wang et al.
(2009). The UV to mid-IR models are standard low $z$ quasar SEDs.
The FIR through radio model is that for an active star forming galaxy. 
Middle: The CO excitation in $z \sim 6$ quasar host galaxies
(Wang etal 2010). Right: The star formation law for nearby and high
$z$ galaxies, including the $z \sim 6$ quasar host galaxies.}
\end{figure}

We have performed extensive studies of the SEDs of the $z \sim 6$
quasars from rest frame UV to radio wavelengths, with reasonable
sampling of the rest frame FIR. Figure 1a shows the results for the
most distant quasar known, J1148+5251 at $z = 6.42$. Through the rest
frame mid-IR the SED is consistent with the standard low $z$ quasar
templates. However, in the FIR, these mm-detected quasars show a clear
excess above the low $z$ quasar template. The FIR excess can be
reasonable fit with dust at 50K. Extrapolating into the radio, we find
that the SEDs of these sources are consistent with the radio--FIR
correlation for star forming galaxies (Yun et al.  2001). Based on
this result, as well as similar circumstantial evidence from molecular
gas and FSL emission below, we conclude that the host galaxies are
undergoing a major starburst episode, co-eval with the AGN. The
implied star formation rates are $\sim 10^3$ M$_\odot$ year$^{-1}$.

\subsection{Molecular gas: the fuel for star formation}

We have searched for CO emission in 8 FIR-lumious $z \sim 6$ quasars
host galaxies, and we detect the molecular gas in all cases (Wang et
al 2010). The implied molecular gas masses are $\sim 10^{10}
({\alpha/0.8})$ M$_\odot$, and line widths range from 200 km s$^{-1}$
to 800 km s$^{-1}$.

We have observations of multiple CO transitions in a number of the 
$z \sim 6$ quasars (Figure 1b). The CO is highly excited, consistent
with constant brightness temperature ($S_\nu \propto \nu^2$) up to 
CO 5-4, at least. Standard radiative transfer modeling (LVG) implies
warm ($\ge 50$K), and dense ($\sim 10^4$ cm$^{-3}$) molecular gas. 
Gas at this density is only seen in the star forming cores of 
GMCs in the Milky Way, on scales $\sim 1$pc. In the case of
the quasars, we find these conditions persist on kpc-scales. 

We have investigated the integrated relationship between gas mass and
star formation for the $z \sim 6$ quasar host galaxies using the
standard 'Kennicutt-Schmidt' law, ie. the relationship between CO
luminosity and FIR luminosity (or surface brightness; Figure 1c)). A
correlation between CO and FIR luminosity has been well established
for nearby galaxies, as well high redshift star forming galaxies (Gao
\& Solomon 2004). The $z \sim 6$ quasar host galaxies fall within the
scatter in this relationship, as defined by the star forming galaxy
populations. The origin of this relationship has been extensively
discussed in the literature. Herein, we make two simple
points. First, the fact that the quasar hosts fall within the
established relationship is further circumstantial evidence for star
formation.  And second, the relationship is non-linear, with a
power-law index of 1.5. This implies a decreasing gas consumption
timescale with increasing luminosity. The quasar hosts would consume
all of their molecular gas in $10^7$ years at the current star
formation rates, as compared to eg. the Milky Way, for which the gas
consumption timescale is 30 times longer.

Interferometric arrays allow for high resolution imaging of the
molecular gas in these distant galaxies. Figure 2a shows the CO
distribution and kinematics for J1148+5251 at $z = 6.42$ at $0.3"$
resolution (Walter et al. 2004). The molecular gas extends to a radius
of at least $\sim 3$kpc.  Higher resolution observations ($0.15"$)
show two high T$_B$ peaks (35K) with sizes $\sim 1$kpc, separated by
$\sim 2$kpc, each comprising about 1/4 of the total emission.  The
velocity field implies a dynamical mass within 3 kpc radius of $5.5
\times 10^{10}$ M$_\odot$. The dynamical mass is within a factor few
of the molecular gas mass, suggesting that the inner few kpc of the
galaxies are baryon-dominated (as is seen in low $z$ ellipticals), and
that the molecular gas comprises a substantial fraction of the
dynamical mass ($\sim 10\%$ to 50\%).

\begin{figure}
  \includegraphics[height=.18\textheight]{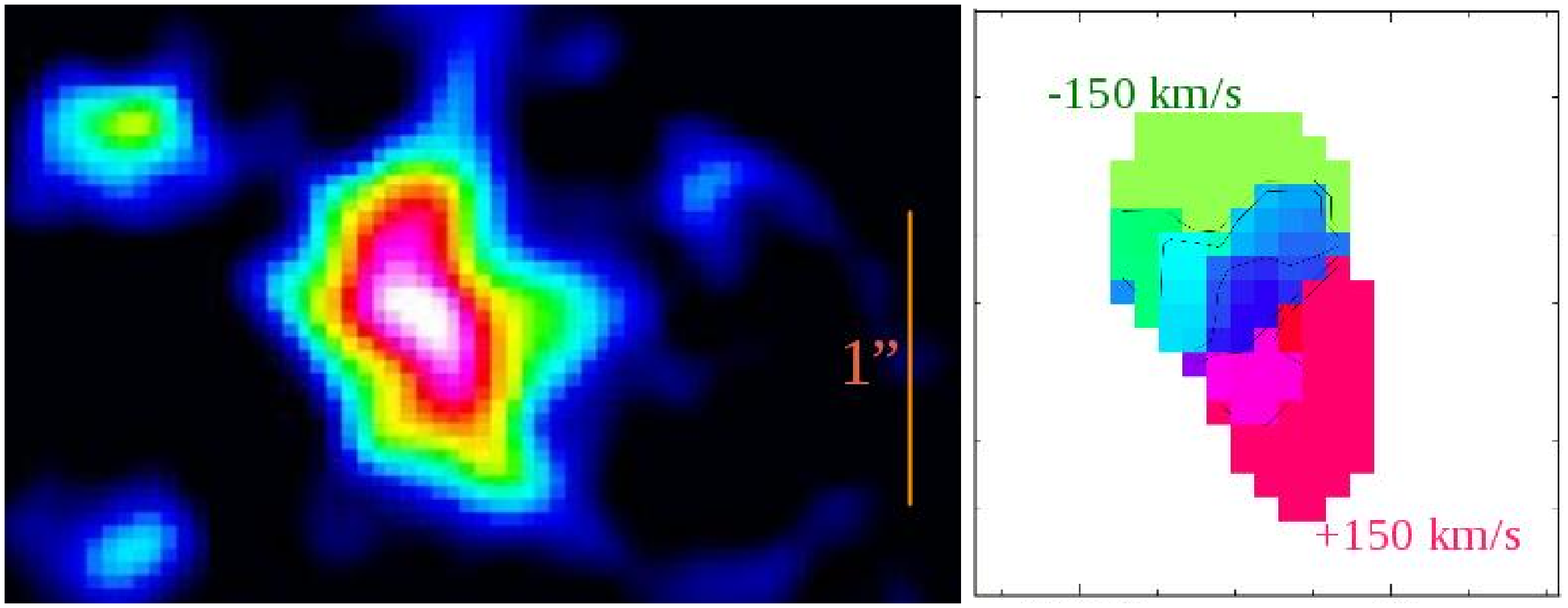}
  \includegraphics[height=.18\textheight]{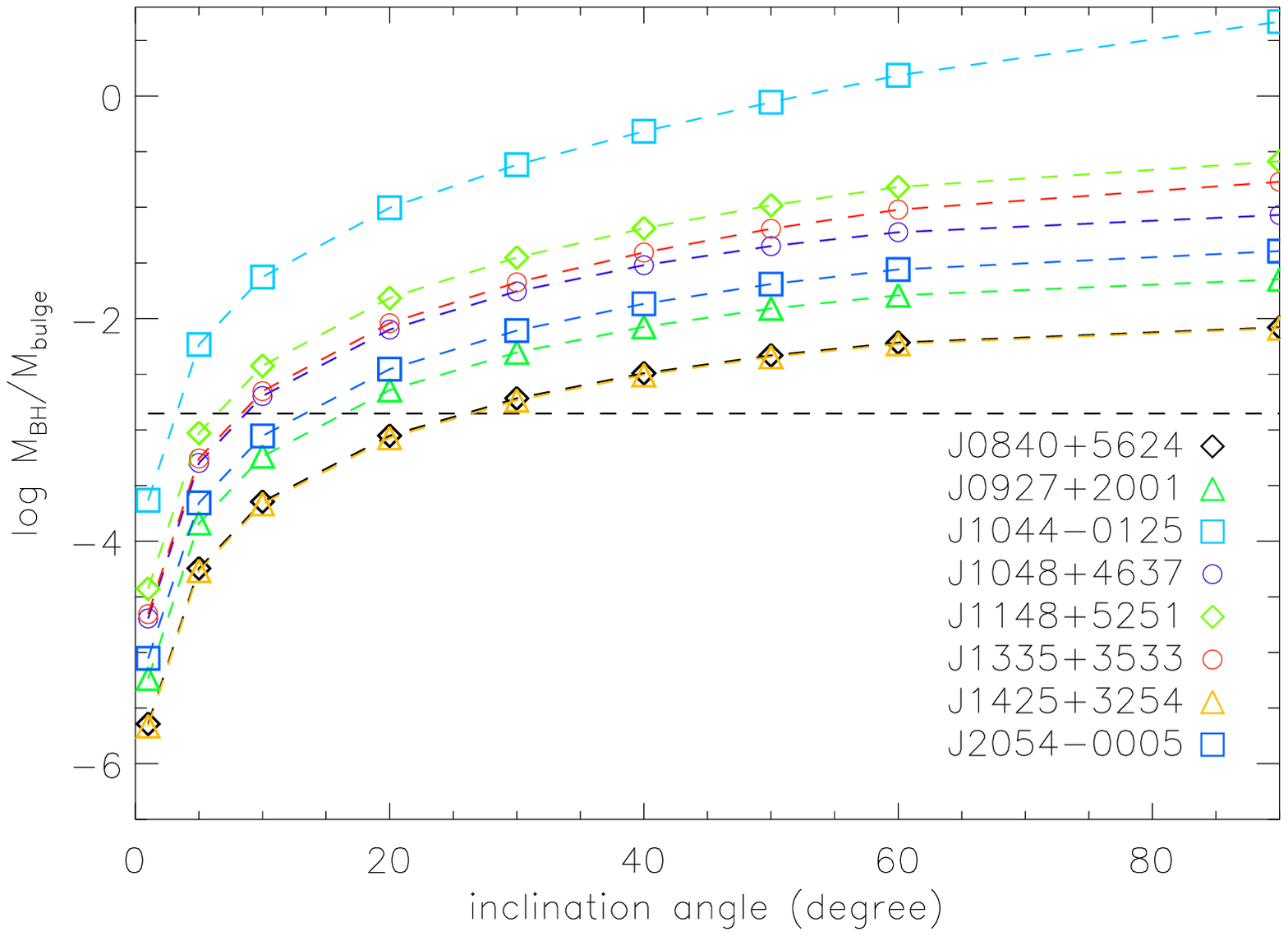}
  \caption{Left: The CO 3-2 distribution  in J1148+5251 
derived from VLA observations (Walter et al. 2004; 2003). 
Center: the CO 7-6 velocity field of J1148+5251 derived
from PdBI observations (Riechers et al. 2010, in prep).
Right: The black hole\ -- bulge mass ratio for
the $z \sim 6$ quasars as a function of assumed inclination
angle for the disk (Wang et al. 2010). The dash line represents
the low $z$ ratio.}
\end{figure}

High resolution imaging of gas dynamics allows for study of the
evolution of the black hole -- bulge mass relation at high
redshift. Indeed, these CO observations present the only method in the
near term to perform such a test in the most distant galaxies. Imaging
of a few $z \ge 4$ quasars shows a systematic departure from the low
$z$ relationship (Riechers et al.  2008; 2009; Walter et
al. 2004). Wang et al. (2010) find that, assuming random inclination
angles for the molecular gas, the $z \sim 6$ quasars are, on average,
a factor 15 away from the black hole -- bulge mass relation, in the
sense of over-massive black holes.  Alternatively, all of the $z \sim 6$
quasars could be close to face-on, with inclination angles relative to
the sky plane all $< 20^o$ (Figure 2c).  High resolution imaging of
the CO emission from these systems is required to address the
interesting possibility that the black holes form before the host
spheroids.

\subsection{Fine structure lines: a maximal starburst disk}

Atomic fine structure line emission, and in particular, the [CII] 158
$\mu$m line, is thought to be the dominant ISM gas cooling line,
tracing the CNM and photon-dominated regions associated with star
formation (Cormier et al. 2010).  These lines require space
observations to be studied in nearby galaxies, however, at high
redshift, the lines redshift into the submm, and hence can be studied
with existing ground-based telescopes. We have started a systematic
search for [CII] emission from $z > 6.2$ quasar host galaxies using
the PdBI. Thus far, we have three detections of the [CII] line,
including J1148+5251 (Figure 3a). One [CII] detection is for a quasar
which is not detected in submm continuum emission. Generally, recent
results on [CII] emission from high $z$ star forming galaxies suggest
a broad scatter in the [CII]/FIR ratio, by about two orders of
magnitude.

Figure 3b shows images of the [CII] emission from J1148+5251 at 0.25$"$
resolution from the PdBI (Walter et al. 2009). The [CII] emission is
extended over about 1.5 kpc. If [CII] traces star formation, the
implied star formation rate per unit area $\sim 10^3$ M$_\odot$
year$^{-1}$ kpc$^{-2}$. This value corresponds to the predicted upper
limit for a 'maximal starburst disk' by Thompson et al. (2005), ie. a
self-gravitating gas disk that is supported by radiation pressure on
dust grains. Such a high star formation rate areal density has been
seen on pc-scales in Galactic GMCs, as well as on 100 pc scales in the
nuclei of nearby ULIRGs. For J1148+5251 the scale for the disk is yet
another order of magnitude larger.

\begin{figure}
  \includegraphics[height=.2\textheight]{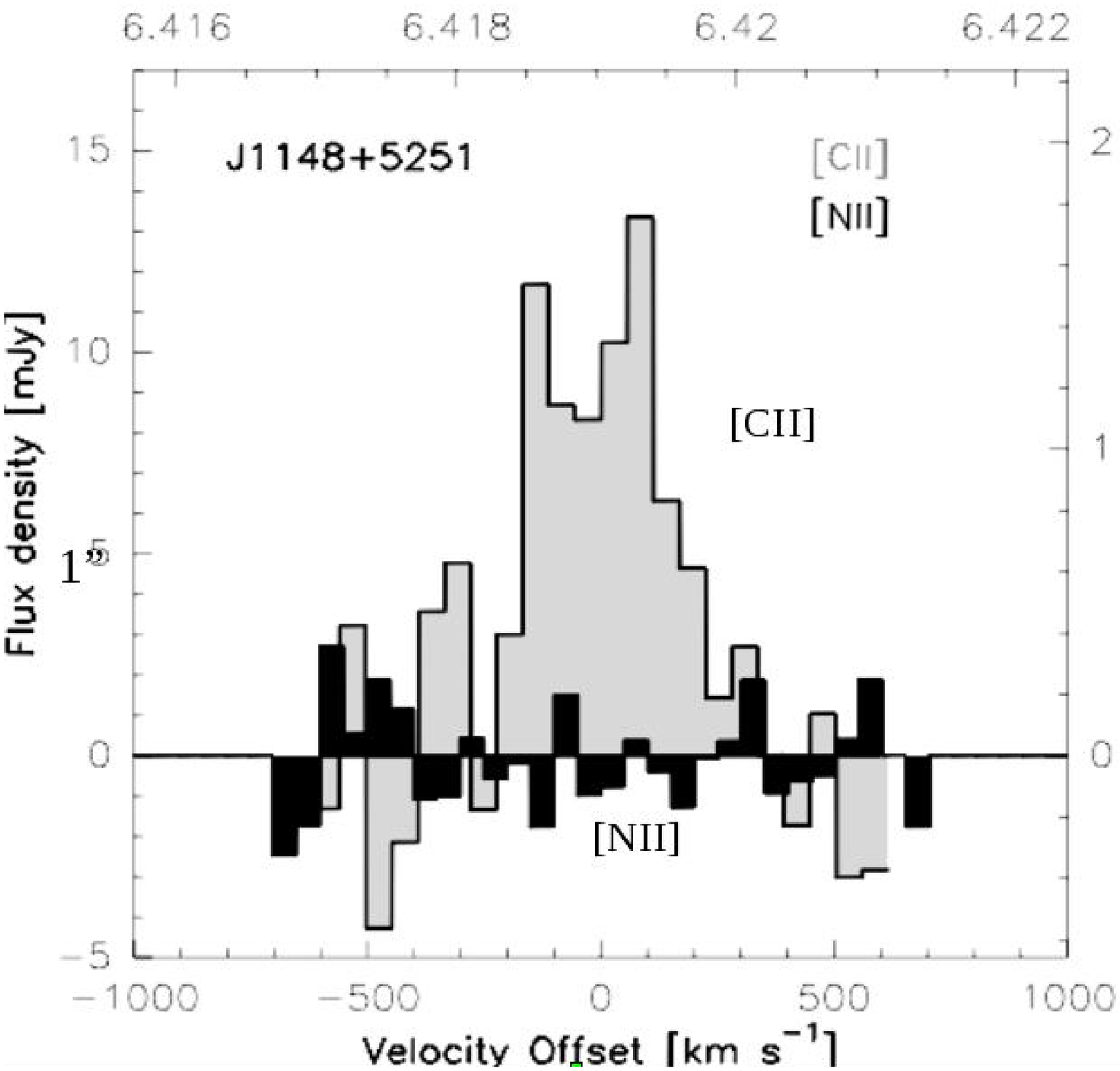}
  \includegraphics[height=.2\textheight]{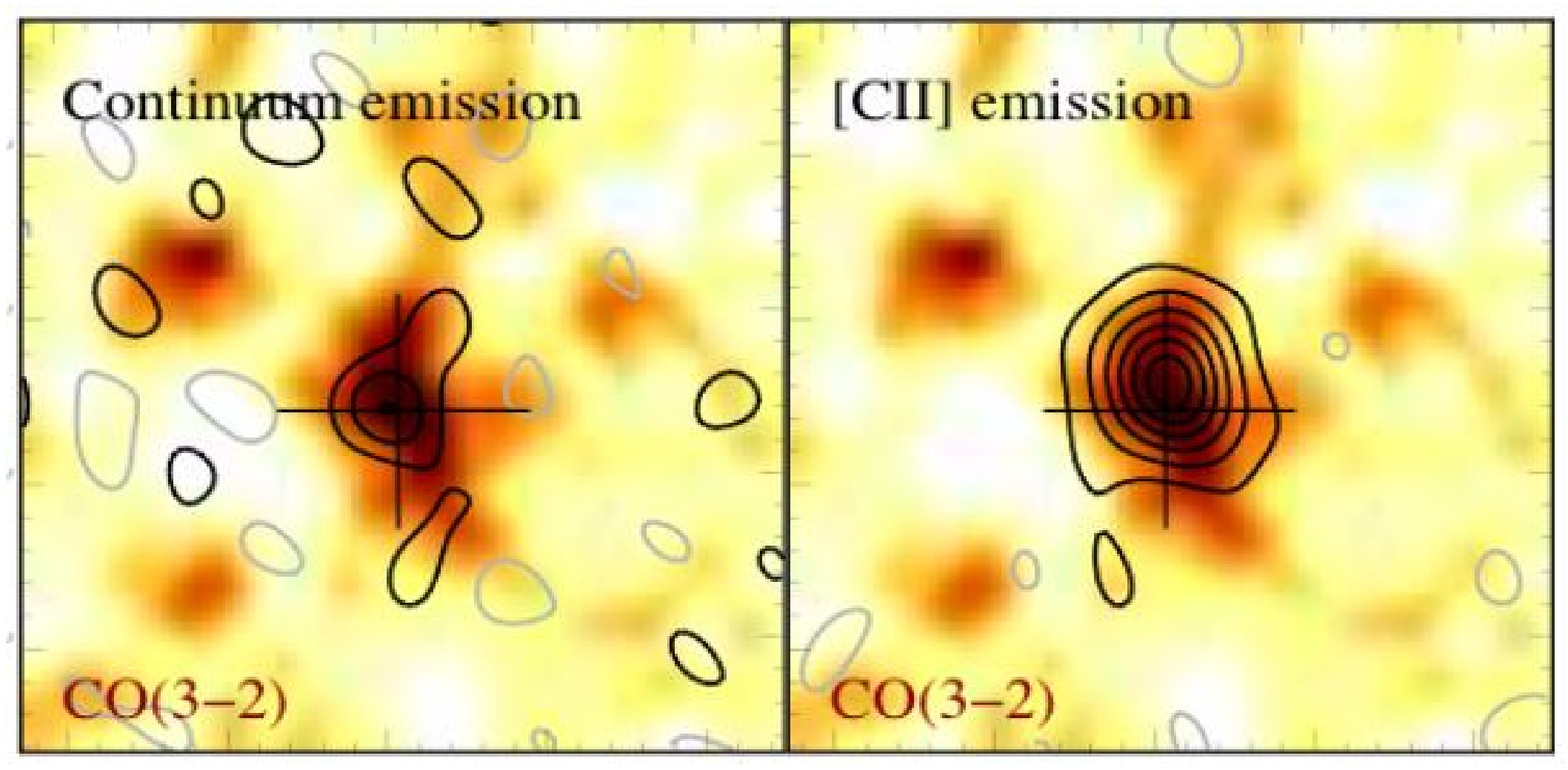}
  \caption{Left: the [CII] 158 $\mu$m spectrum of J1148+5251, 
plus an upper limit for [NII] 205$\mu$m from the PdBI (Maiolino et al.
(2005). Right: 
PdBI imaging of the dust and [CII] emission (contours, left and right,
respectively) at $0.25"$ resolution,
plus the VLA CO 3-2 in greyscale (Walter et al. 2009). }
\end{figure}

\subsection{Quasar near-zones}

Fan et al. (2006) present evidence for ionized regions, or quasar
near-zones (NZ), surrounding $z \sim 6$ quasars on Mpc-scales,
presumably caused by radiation from the quasars ionizing the local
(partially neutral) IGM. The evidence is in the form of excess
emission in the wing of the Ly$\alpha$ line leaking to lower redshift
due to the ionization of the immediate environs of quasar. Their
analysis suggests a decrease in the size of these spheres from $z =
5.7$ to 6.4, qualitatively consistent with an increase in the neutral
fraction of the IGM over this narrow redshift range. However, the Fan
et al. study had inaccurate host galaxy redshifts (based,
predominantly, on UV emission lines), and had a limited sample size.

We have improved on the analysis using a sample of 27 quasars between
$z=5.7$ to 6.4, include more sources than previous studies, and more
accurate redshifts for the host galaxies, with 8 CO molecular line
redshifts and 9 MgII redshifts.  We confirm the trend for an increase
in NZ size with decreasing redshift, with the luminosity normalized
proper size evolving as: $\rm R_{NZ,corrected} = (7.4 \pm 0.3) - (8.0
\pm 1.1) \times (z-6)$ Mpc.  While derivation of the absolute IGM
neutral fraction remains difficult with this technique, the evolution
of the NZ sizes suggests a decrease in the neutral fraction of
intergalactic hydrogen by a factor $\sim 9.4$ from $z=6.4$ to 5.7, in
its simplest interpretation.  Alternatively, recent numerical
simulations suggest that this rapid increase in near-zone size from
$z=6.4$ to 5.7 is due to the rapid increase in the background
photo-ionization rate at the end of the percolation or overlap phase
of cosmic reionization, when the average mean free path of ionizing
photons increases dramatically (Bolton et al. 2010).  In either case,
the results are consistent with the idea that $z \sim 6$ to 7
corresponds to the tail end of cosmic reionization. The scatter in the
normalized NZ sizes is larger than expected simply from measurement
errors, and likely reflects intrinsic differences in the quasars or
their environments. We find that the near-zone sizes increase with
quasar UV luminosity, as expected for photo-ionization by
quasar radiation.

\begin{figure}
  \includegraphics[height=.25\textheight]{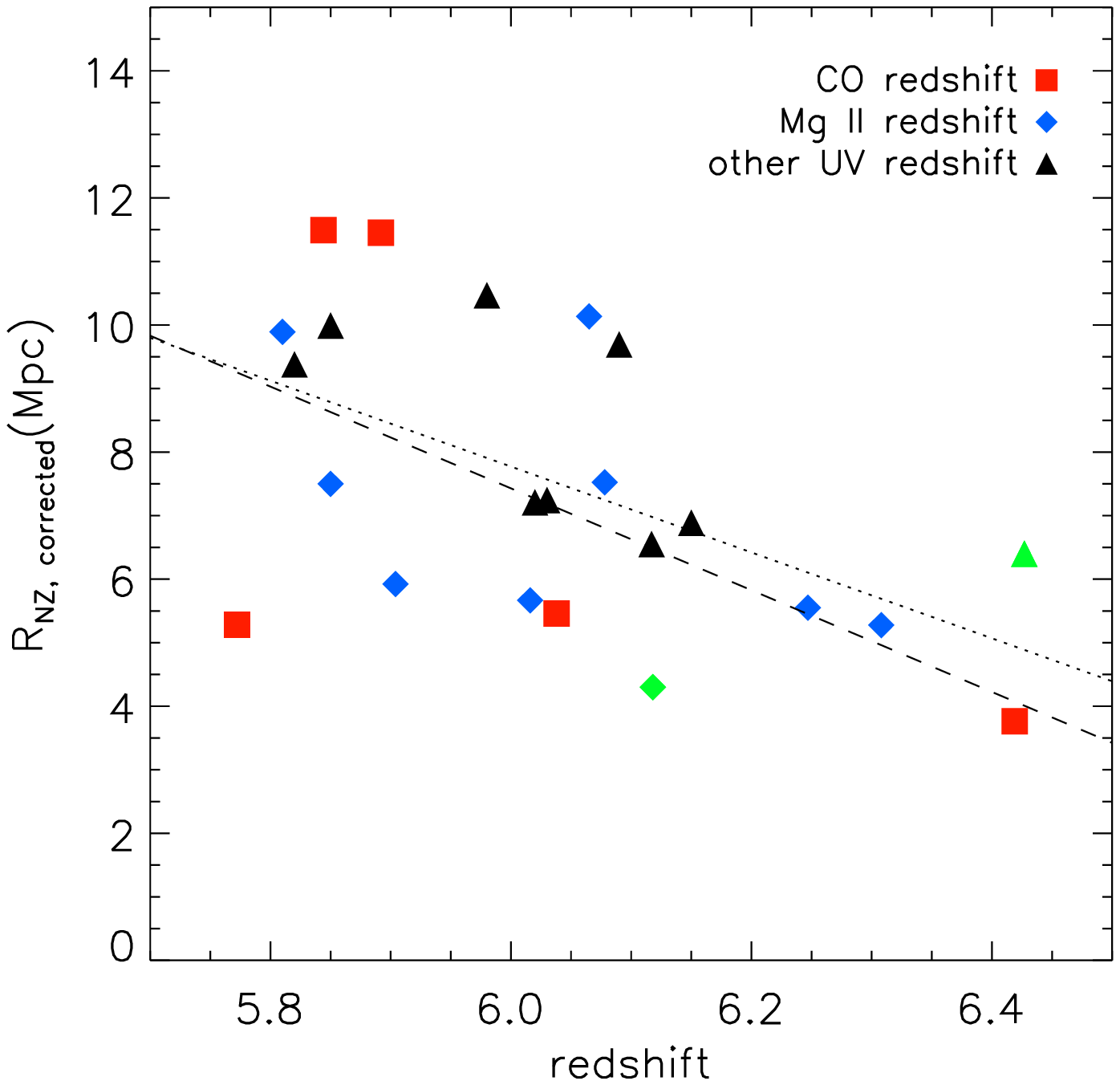}
  \includegraphics[height=.25\textheight]{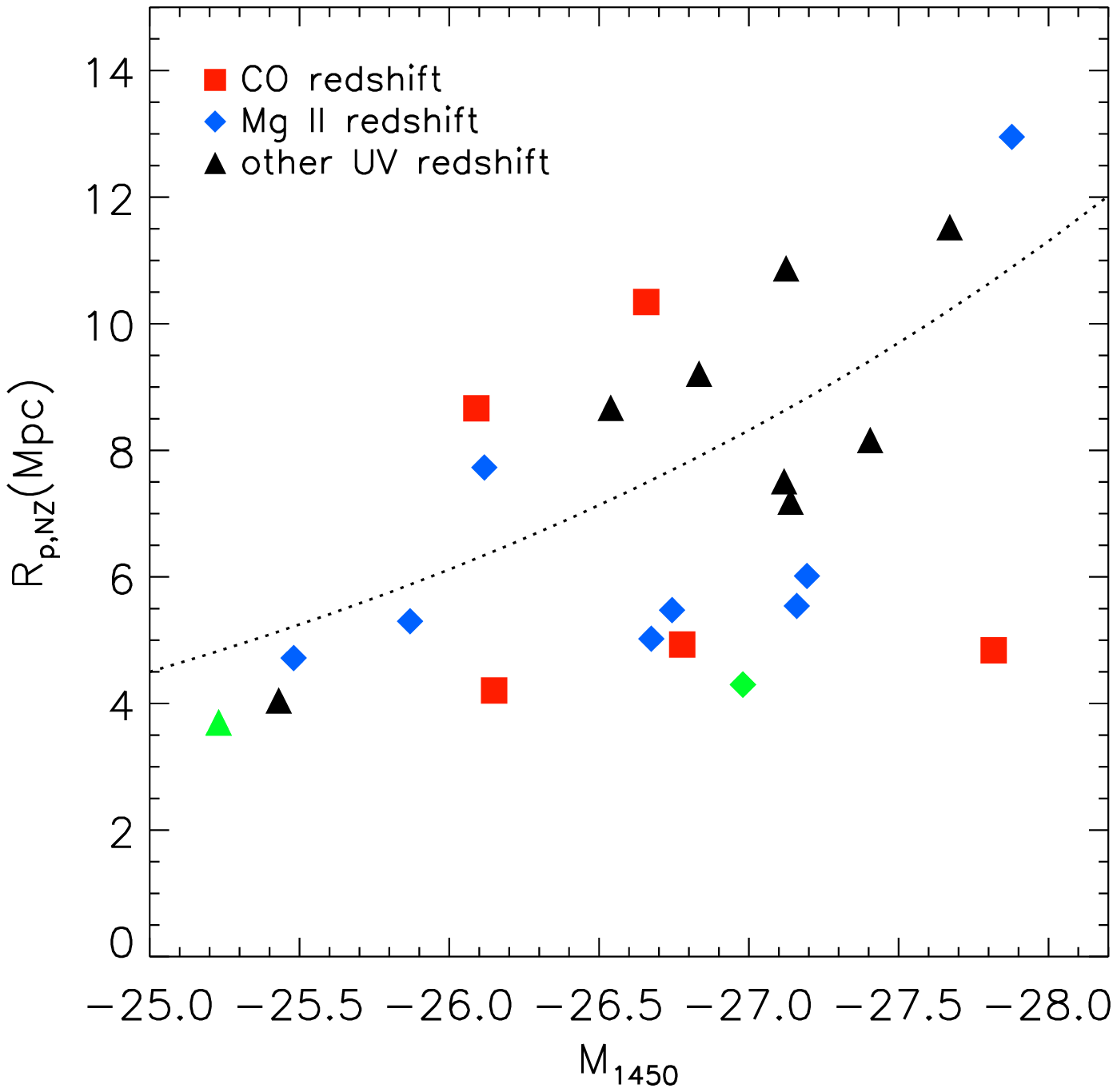}
  \caption{Left: Radii of the luminosity normalized 
quasar near zones  versus redshift (Carilli et al. 2010). 
The long dash line show a weighted linear fit to the
data with $\rm R_{NZ,corrected} = (7.4 \pm 0.3) - (8.0 \pm 1.1) \times
(z-6)$.  Right: The relationship between $\rm R_{\rm p,NZ}$ and $M_{1450}$. 
The dotted line is not a fit to the data, but shows the
relationship: $\rm R_{NZ} \propto \dot{N}_Q^{1/3} \propto 10^{-M_{1450}/7.5}$, 
as expected for photo-ionization by the quasars.
}
\end{figure}

\subsection{Conclusions: ALMA and EVLA}

Overall, the study of $z \sim 6$ quasars suggest massive starbursts in
the host galaxies of the FIR luminous sample, ie. co-eval formation of
a massive galaxy and a supermassive black hole within 1 Gyr of the Big
Bang. Li et al. (2007) have investigated this possibility using their
large volume cosmic structure formation simulation. They select the
most massive halo in the 3 cGpc$^3$ volume, and follow the formation
of the galaxy and supermassive black hole, starting at $z = 14$.  They
find that a massive galaxy can form by $z \sim 6$ through a series of
gas rich mergers, driving the formation rates over 10$^3$ M$_\odot$
year$^{-1}$ at times. The 10$^9$ M$_\odot$ black hole also forms via
Eddington limited accretion and black hole mergers. The system should
eventually evolve into a giant elliptical galaxy at the center of a
massive cluster of galaxies. One caveat is that that the duty-cycle
must be high, ie. the galaxy has been active for much of the time
since $z = 14$. Further key observations to test these ideas include:
(i) a search for the expected companion galaxies in the clusters, and
(ii) observations of the stellar population of the host galaxy.

Our studies demonstrate the power of radio observations to study
the dust, gas, and (obscuration-free) star formation in the first
galaxies. However, we are currently limited to the most active,
massive galaxies at these high redshifts. Fortunately, the EVLA and
ALMA are both close to completion, opening a new window into the the
study of the first galaxies. Figure 7 shows the capabilties for
studying the line emission from distant star forming galaxies. The
EVLA, with its sensitivity and broad fractional bandwidth, will allow
for large cosmic volume, blind surveys for molecular gas in early
galaxies. Likewise, the two orders of magnitude improvement in submm
sensitivity with ALMA enables the study of the fine structure lines
and dust in normal star forming galaxies (SFR $\sim 1$ to 10 M$_\odot$
year$^{-1}$) well into cosmic reiozination ($z \sim 6$ to 10).

ALMA and the EVLA represent an order of magnitude, or more,
improvement in most observational capabilities from 1 GHz to 1
THz. These facilities provide a powerful new tool to study the 'other
half of galaxy formation.'

\begin{figure}
  \includegraphics[height=.25\textheight]{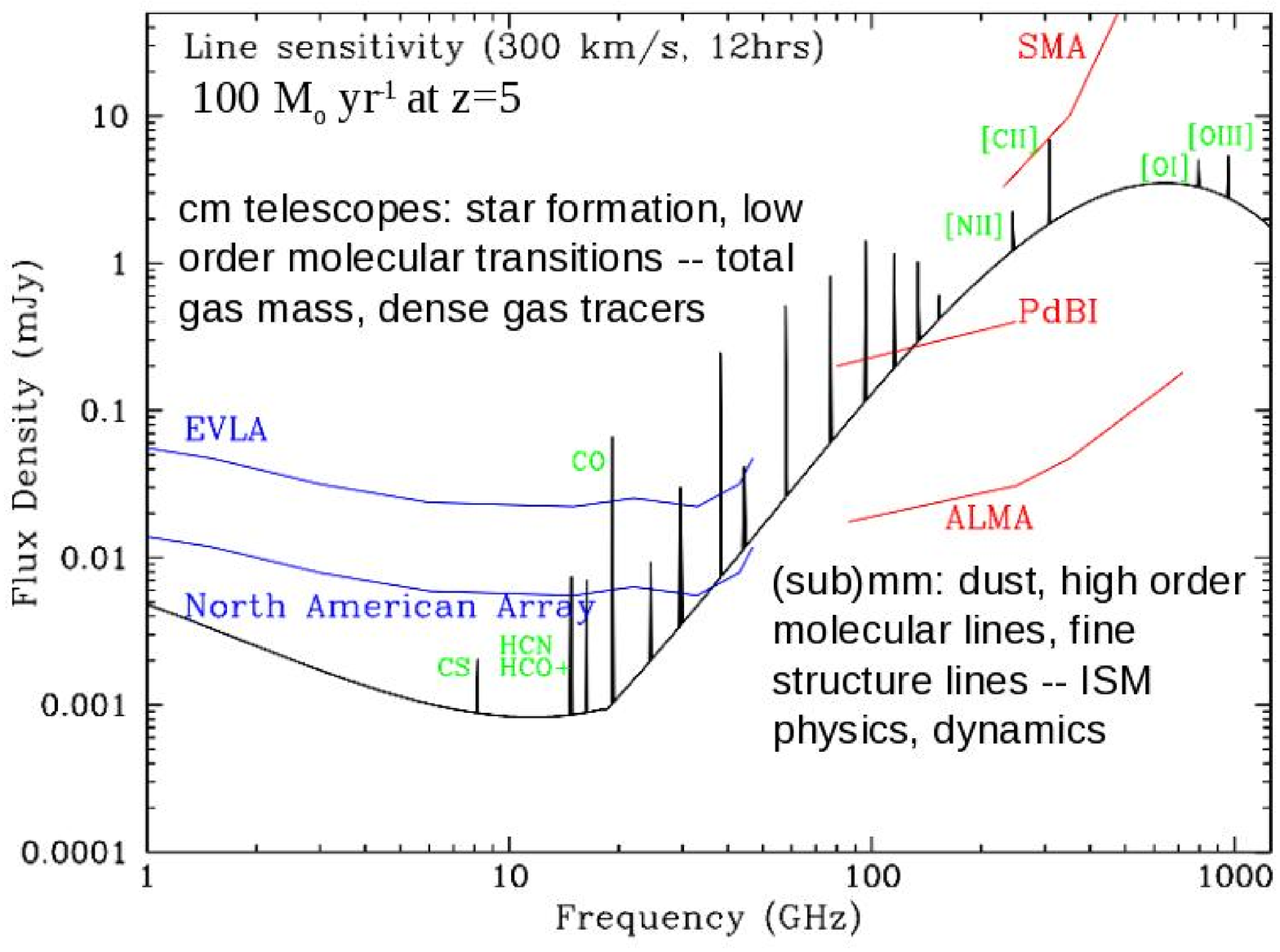}
  \includegraphics[height=.25\textheight]{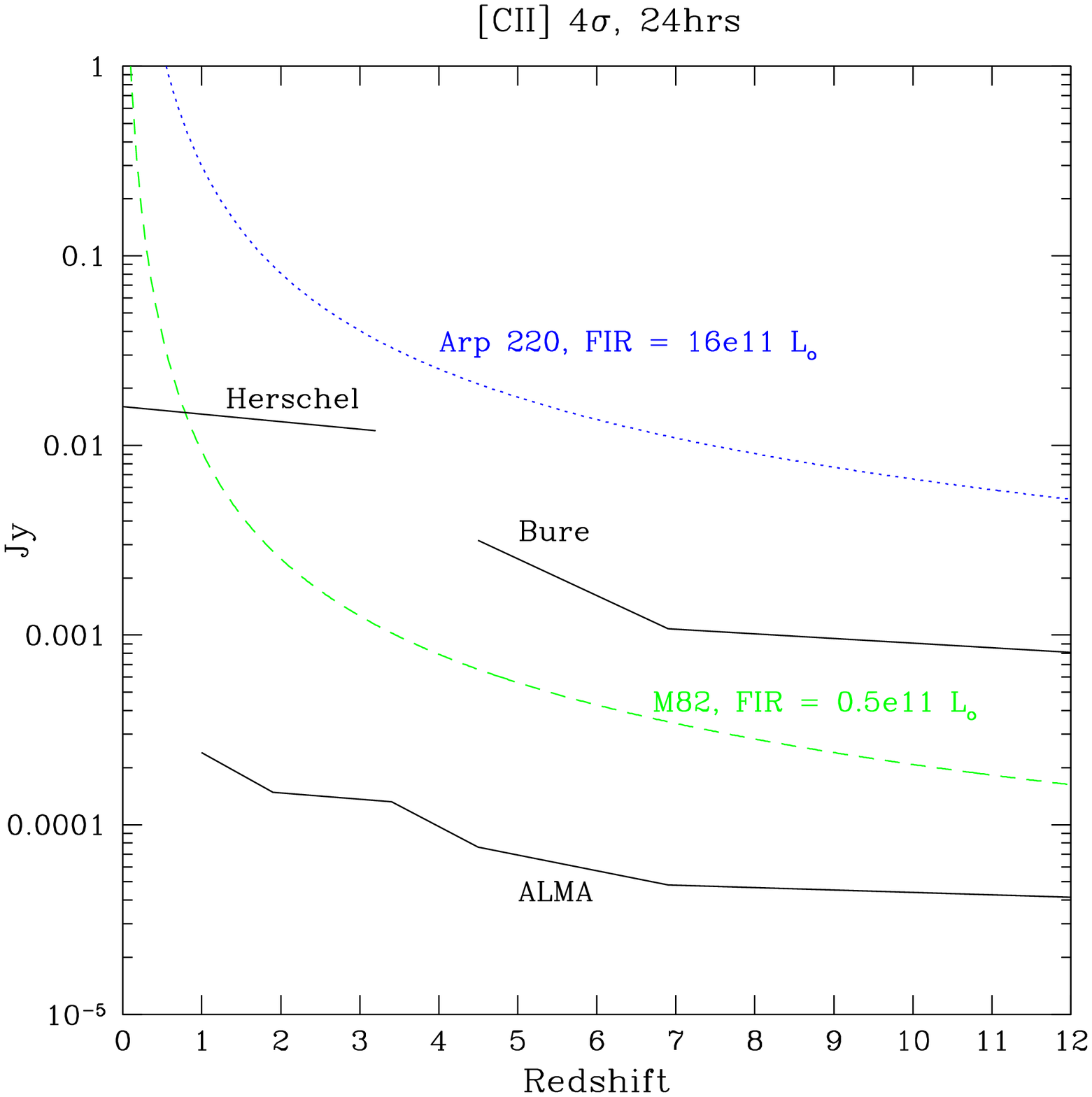}
  \caption{Left: the spectrum of an active star forming galaxy at $z = 5$
plus the sensitivity of the current and future interferometers. Right: 
the expected [CII] line peak flux density versus redshift for active and
dwarf star forming galaxies, plus the sensitivity of ALMA and other instruments.}
\end{figure}

\vskip 0.1in

\small{Bolton, J.S. et al.  \emph{MNRAS}, in press (2010)}

\small{Carilli, C.L. et al.  \emph{ApJ} \textbf{714}, 834 (2010)}

\small{Cormier et al.  \emph{A\& A} in press (2010)}

\small{Draine, B.  \emph{ARAA} \textbf{41}, 241 (2003)}

\small{Dwek, E. et al.  \emph{ApJ} \textbf{662}, 927 (2007)}

\small{Elvis, M. et al. \emph{ApJ} \textbf{567}, L107 (2002)}

\small{Fan, X. et al.  \emph{AJ} \textbf{632}, 117 (2006)}

\small{Gao, Y. \& Solomon, P.  \emph{ApJ} \textbf{606}, 271 (2004)}

\small{H\"aring, N. \& Rix, W.  \emph{ApJ} \textbf{604}, L89 (2004)}

\small{Li, Y. et al.  \emph{ApJ} \textbf{665}, L187 (2007) }

\small{Maiolino, R. et al.  \emph{A\& A} \textbf{40}, L51 (2005)}

\small{Perley, D. et al.  \emph{ApJ} in press (2010)}

\small{Riechers, D. et al. \emph{ApJ} \textbf{690}, 463 (2009) }

\small{Riechers, D. et al. \emph{ApJ} \textbf{686}, L9 (2008)}

\small{Stratta, G. et al.  \emph{ApJ} \textbf{661}, L9 (2007)}

\small{Thompson, T. et al.  \emph{ApJ} \textbf{630}, 167 (2005)}

\small{Venkatesan, A. et al.  \emph{ApJ} \textbf{640}, 31 (2006)}

\small{Walter, F. et al. \emph{Nature} \textbf{457}, 699 (2009)}

\small{Walter, F et al. \emph{ApJ} \textbf{615}, L17 (2004)}

\small{Walter, F. et al. \emph{Nature} \textbf{424}, 406 (2003)}

\small{Wang, R. et al. \emph{ApJ} \textbf{714}, 699 (2010)}

\small{Wang, R. et al. \emph{ApJ} \textbf{687}, 848 (2008)}

\small{Willott, C. et al.  \emph{AJ} \textbf{139}, 906 (2010)}

\small{Yun, M.S. et al.  \emph{ApJ} \textbf{554}, 803 (2001)}

\end{document}